%% file: main.tex
\documentclass[fleqn,10pt, twocolumn]{olplainarticle}

\newgeometry{vmargin=25mm, hmargin={20mm,20mm}} 

\input{macros}

\usepackage{pdfpages}
\pdfoutput=1

\def\appendixname{Appendix }
\renewcommand\appendix{\par
  \setcounter{section}{0}%
  \setcounter{subsection}{0}%
  \setcounter{equation}{0}
  \setcounter{figure}{0}
  \gdef\thefigure{\Alph{section}.\arabic{figure}}%
  \gdef\thetable{\Alph{section}.\arabic{table}}%
  \gdef\thesection{\appendixname~\Alph{section}}%
  \gdef\theequation{\Alph{section}.\arabic{equation}}%
  \addtocontents{toc}{\let\string\numberline\string\tmptocnumberline}{}{}
}

\newdimen\appnamewidth
\def\tmptocnumberline#1{%
   \setbox0=\hbox{\appendixname}
   \appnamewidth=\wd0
   \addtolength\appnamewidth{2.5pc}
   {#1}
}

\title{FlowNet-PET: Unsupervised Learning to Perform Respiratory Motion Correction in Positron Emission Tomography}

\author[1]{Teaghan O'Briain}
\author[2,3]{Carlos Uribe}
\author[4]{Kwang Moo Yi}
\author[5,6]{Jonas Teuwen}
\author[5,7]{Ioannis Sechopoulos}
\author[1]{Magdalena Bazalova-Carter}

\affil[1]{Department of Physics and Astronomy, University of Victoria, Victoria, BC Canada V8W 3P2}
\affil[2]{Functional Imaging, BC Cancer, Vancouver, Canada V5Z 1G1}
\affil[3]{Department of Radiology, University of British Columbia, Vancouver, BC Canada V6T 2B5}
\affil[4]{Department of Computer Science, University of British Columbia, Vancouver, BC Canada V6T 1Z4}
\affil[5]{Department of Medical Imaging, Radboud University Medical Center, Nijmegen, Netherlands 6525 GA}
\affil[6]{Netherlands Cancer Institute Centre, Amsterdam, Netherlands 1066 CX}
\affil[7]{Technical Medical Centre, University of Twente, Enschede, Netherlands 7522 NH}


\input{sections/0_Abstract.tex}

\begin{document}

\flushbottom
\maketitle
\thispagestyle{empty}

\input{sections/1_Introduction}
\input{sections/2_Methods}
\input{sections/3_Experiments}
\input{sections/4_Results}
\input{sections/5_Discussion}
\input{sections/6_Conclusion}
\input{sections/7_Appendix}

\section*{Acknowledgments}

The authors would like to thank Paul Segars from Duke University; in addition to providing ongoing support for the XCAT phantom, Paul graciously allowed for the training and tests sets to be made publicly available. We would also like to express our gratitude to Erik Aarntzen who provided the clinical activity distributions that we used to define the range of activities for the digital phantoms. This research was enabled in part by the support and resources provided by WestGrid (www.westgrid.ca) and Compute Canada (www.computecanada.ca). This work was supported in part by an NSERC CGS-M, a Discovery Grant, and the Canada Research Chairs Program.

\bibliography{refs}

\end{document}

%% file: macros.tex
\usepackage{afterpage}

\newcommand{\loss}{\mathcal{L}}
\newcommand{\figref}[1]{Figure~\ref{#1}}

\newcommand{\tabref}[1]{Table~\ref{#1}}
\newcommand{\secref}[1]{Section~\ref{sec:#1}}
\newcommand{\appenref}[1]{\ref{sec:#1}}

\newcommand{\fnp}[0]{\textsc{FlowNet-PET }}
\newcommand{\fnpp}[0]{\textsc{FlowNet-PET}}

%% file: sections/0_Abstract.tex
\begin{abstract}
\noindent {\bf Background:} Accounting for respiratory motion in PET is often accomplished using retrospective phase binning. Due to the inefficient use of the acquired data, this technique necessitates protocol adjustments in the form of either longer scan times or higher injected activities to produce diagnostic quality images.\\ 
{\bf Purpose:} To efficiently correct for respiratory motion in PET imaging by developing an interpretable and unsupervised deep learning technique called \fnpp. \\
{\bf Methods:} The network was trained to predict the optical flow between two PET frames from different breathing amplitude ranges. The trained model aligns different retrospectively-gated PET images, providing a final image with similar counting statistics as a non-gated image, but without the blurring effects. \fnp was applied to anthropomorphic digital phantom data, which provided the possibility to design robust metrics to quantify the corrections. When comparing the predicted optical flows to the ground truths, the median absolute error was found to be smaller than the pixel and slice widths. The improvements were illustrated by comparing against images without motion and computing the intersection over union (IoU) of the tumors as well as the enclosed activity and coefficient of variation (CoV) within the no-motion tumor volume before and after the corrections were applied.\\
{\bf Results:} The average relative improvements provided by the network were 64\%, 89\%, and 75\% for the IoU, total activity, and CoV, respectively. \fnp achieved similar results as the conventional retrospective phase binning approach, but only required one sixth of the scan duration. \\
{\bf Conclusions:} \fnp accurately predicts interpretable optical flows that correct for the respiratory motion found in PET data. By using unsupervised learning, this method can be readily transferred to clinical data as well as other imaging modalities. The code and data have been made publicly available (https://github.com/teaghan/FlowNet\_PET). \\
\end{abstract}

%% file: sections/1_Introduction.tex
\section{Introduction}
\label{sec:introduction}

Positron emission tomography (PET) is used to measure the distribution of a radionuclide administered to the patient, which provides an assessment of the physiological processes associated with the uptake of that radionuclide. When a cancer-targeting radiopharmaceutical (such as fluorodeoxyglucose (FDG) labeled with $^{18}$F) is chosen, PET can be used to identify cancer cells with high metabolic activity \citep{wahl_2008}. However, PET is limited by the rate of uptake of the radiopharmaceutical in the tissue and by the radioactive decay of the radionuclide involved, which make long acquisition times (2-4 min/bed position in clinical routine) necessary to produce diagnostic quality images. Due to patient motion during the acquisition, PET images represent the motion-averaged activity distribution over several minutes, and therefore, motion-induced artifacts are common \citep{Osman2003}. This is especially true when imaging areas that are more prone to movement due to breathing, such as those in the lower thorax and upper abdomen. The end result is poorer image quality and a reduction in apparent uptake, which can impair diagnostics and cancer localization for patients with lung cancer, esophageal cancer, pancreatic cancer, and liver lesions \citep{Grootjans2015, vanderVos2014, Van_der_Vos2015}.

To account for motion, retrospective phase binning (RPB) is commonly used to produce respiratory-gated PET images \citep{Didierlaurent2012}. In this scenario, the patient’s breathing pattern is monitored using an external device \citep{Nehmeh2002, Nehmeh2003, Erdi2004} or data-driven methods \citep{Schleyer2009, Buther2010, Kesner2010} and the events detected during a single phase of the breathing motion are used to reconstruct the image while the remaining detections are disregarded. Although this mitigates the majority of the breathing motion, the main disadvantage of this technique is the poor efficiency of the data acquisition, which results in noisy images. Solving this issue often requires a higher injected activity, resulting in a higher patient effective dose and imaging costs. Alternatively, the acquisition times can be increased (6 min/bed position), which decreases patient throughput and increases patient discomfort. Neither of these solutions are ideal.

Alternatively, registration between the different phases of the breathing motion using a global non-rigid registration method have been proposed \citep{qfreeze}. This has the advantage that all of the detected events are used to generate a single image. Unfortunately, registration of noisy images (i.e., individual gated ones) can be challenging. As a result, these methods often depend on the registration of matching gated low-dose computed tomography images, which can be misaligned with the gated PET images \citep{Moller188}. Furthermore, the details of these methods have not been made publicly available.

Registration of PET images from different phases or amplitude ranges of the breathing cycle can also be framed as a computer vision problem similar to learning the pixel-wise shift (or \textit{optical flow}) between subsequent frames in a video. Optical flow estimation can be accomplished through supervised learning with convolutional neural networks (CNNs) \citep{Fischer2015, Teed2020}. In this case, the ground truth optical flows between all of the training samples are known \textit{a priori}, which provides the network with a direct reference to compare the predictions against. Since access to ground truth optical flows is not realistic for many applications, unsupervised methods have also been developed \citep{yu2016}. Accessing ground truth optical flows for clinical PET data is impossible, and therefore, unsupervised techniques are required to train a network on patient data.

In this paper, we propose \fnpp, which is a modification of these prior works in computer vision and is capable of performing the registration between three dimensional (3D) PET images acquired during different amplitude ranges of the breathing motion. Once trained, the model groups a set of gated PET images into a motion-corrected single bin, providing a final image without the blurring effects that were initially observed. The method was applied to simulated data; therefore, the work described in this paper reflects a proof-of-concept for the proposed framework. Importantly, due to the unsupervised nature of the training, \fnp can be easily extended to clinical patient images in future work. Additionally, having a neural network estimate the optical flow rather than apply the transformation itself provides a more interpretable model, which is preferable in clinical applications. To improve the accessibility of this method, the code has been made publicly available along with the training and test sets (https://github.com/teaghan/FlowNet\_PET).

%% file: sections/2_Methods.tex
\section{Materials and Methods}
\label{sec:methods}

\subsection{The Training Data}
\label{sec:methods_data}

To facilitate this proof-of-concept, a lung cancer dataset of 300 phantoms was generated using the 4D extended cardiac-torso (XCAT) anthropomorphic digital phantom \citep{Segars2010} with (2$\times$4$\times$4) mm$^3$ voxels. The parameters for each XCAT phantom were varied randomly. These included the gender, axial sections included in the field of view, transaxial shifts, scaling factors in each direction, size and location of the lung lesion, extent of the diaphragm motion, and extent of the chest expansion. The distributions of each of these randomized parameters are shown in \tabref{table:xcat_params}. Additionally, the activity concentration in each organ and tumor was varied based on activity distributions observed in clinical data. 

\input{figures/fig1}

\begin{table}[h]
\setlength\belowcaptionskip{0pt}
\centering
\caption{The distributions of the randomized XCAT parameters used to generate the training data.
\label{table:xcat_params}}
\begin{tabular}{|c|c|}
\hline
Parameter                 & Distribution             \\ \hline
Gender                 &  Male / Female          \\
Axial section in image                  & Any thoracic FOV           \\
Transaxial shifts         & Uniform(-4 cm, 4 cm)     \\
Axial scaling factor      & Uniform(0.7, 1.3)       \\
Transaxial scaling factor & Uniform(0.8, 1.2)       \\
Lesion diameter           & Uniform(1 cm, 3 cm)      \\
Lesion location           & Within the lungs         \\
Axial diaphragm motion          & Uniform(0.9 cm, 2.1 cm)  \\
Chest expansion           & 0.7 $\times$ (Diaphragm motion) \\ \hline
\end{tabular}
\end{table}

The dataset of 300 XCAT phantoms was split into training (270) and validation (30) sets (the test sets are explained in \secref{experiments}). For each phantom, ten frames spaced across a single breathing cycle were created, each frame having (108$\times$152$\times$152) pixels representing the activity distribution throughout the phantom. To emulate a simple representation of PET data acquisition, these distributions were used to sample a random number (between $1\times10^6$ and $9\times10^6$) of counts per frame. Since each phantom consisted of ten frames, this provided 90 paired PET samples to train on per phantom and a total of 24300 training samples.

The intention of creating a large training set of digital phantoms with variability was to provide a simple modeling of some of the variations that could be seen in a dataset of clinical patient data; the goal was not to necessarily recreate real data perfectly. This is because the method is unsupervised and can be retrained (or fine-tuned) on a dataset of clinical gated patient images, which would intrinsically include the necessary variability. However, the XCAT data provides a valuable dataset for the development and initial evaluation of the \fnp motion correction method for several reasons: (1) although the training was unsupervised, having access to the ground truth optical flows provided a robust source of validation when comparing models and reporting the accuracy of the final model, (2) knowing the true activity distribution within the phantom made it possible to track the pixels that belonged to different parts of the phantom (for instance, which pixels belonged to the tumor before and after the corrections were applied), and (3) the same phantom could be used to produce images with and without motion, providing a ground truth image to compare the corrected images against.

\subsection{The \fnp Framework}
\label{sec:methods_flownet}

\fnp was constructed to accomplish the task of predicting the optical flow between two PET images acquired from different amplitude bins of the breathing motion. These two frames are referred to as the input and target PET frames. The estimated flow can then be applied to the input frame through a grid sampling procedure to align the detected counts with the target PET frame. This procedure is outlined in \figref{fig1}.

An in-depth description of the CNN architecture used in \fnp is provided in \appenref{appendix_architecture}, which includes a detailed diagram of the network (\figref{fig10}). In brief, the network takes two PET frames as inputs and subsequently downsamples the images into a feature map through a series of 3D convolutions. The feature map is then upsampled in steps back to the original resolution of the images. At each step, the optical flow is predicted - then enhanced - until an optical flow of the original resolution is produced. The grid sampling procedure then uses the optical flow to index pixels in the input PET frame and place them in new locations within the shifted PET frame. A more detailed explanation on the grid sampling is provided in \appenref{appendix_gridsample}. The method was implemented using Python 3.7 and PyTorch 1.10.0.

\subsection{Training \fnp}
\label{sec:methods_training}

To make the method transferable to clinical patient data, an unsupervised procedure was adopted for training \citep{yu2016}. Namely, rather than comparing the predicted optical flows to the ground truths, a pixel-wise comparison was made between the target frame and the shifted input frame. This objective function is referred to as the photometric loss term, which is explained in detail in \appenref{appendix_training}. Due to the noise found in PET images, performing a pixel-wise comparison could result in the alignment of the noise characteristics rather than the general distribution of the PET detections. Therefore, the unsupervised learning procedure was adjusted slightly such that the target and shifted frames were blurred with a Gaussian filter before being compared in the loss function. Upon investigation, a filter size of (15$\times$15$\times$15) voxels with a standard deviation of (1.6$\times$3.2$\times$3.2) mm was found to produce the best results for this blurring. By minimizing this photometric loss, the network learned to produce optical flows that successfully shifted the signal in the input frame to match the target frame.

Another development that is unique to our method is how we produced invertible optical flows. In PET images, the pixel intensities can be related to the number of detected coincidences in a particular location. Allowing the network to artificially change the number of counts would remove the quantitative imaging advantage of PET. Therefore, to restrict the network to produce optical flows that do not remove or multiply individual counts, we added a second loss term that constrained the network to produce one-to-one pixel correspondences. In short, the optical flows predicted by the network were constrained to be invertible by predicting the flow between the two frames in both directions, where one flow should reverse the effect of the other. The final loss function that was minimized throughout training was a combination of the photometric and invertibility loss terms. More details regarding the formalism of the invertibility constraint are provided in \appenref{appendix_training}.

%% file: figures/fig1.tex
\begin{figure*}[ht]
\centering 
\includegraphics[width=\linewidth]{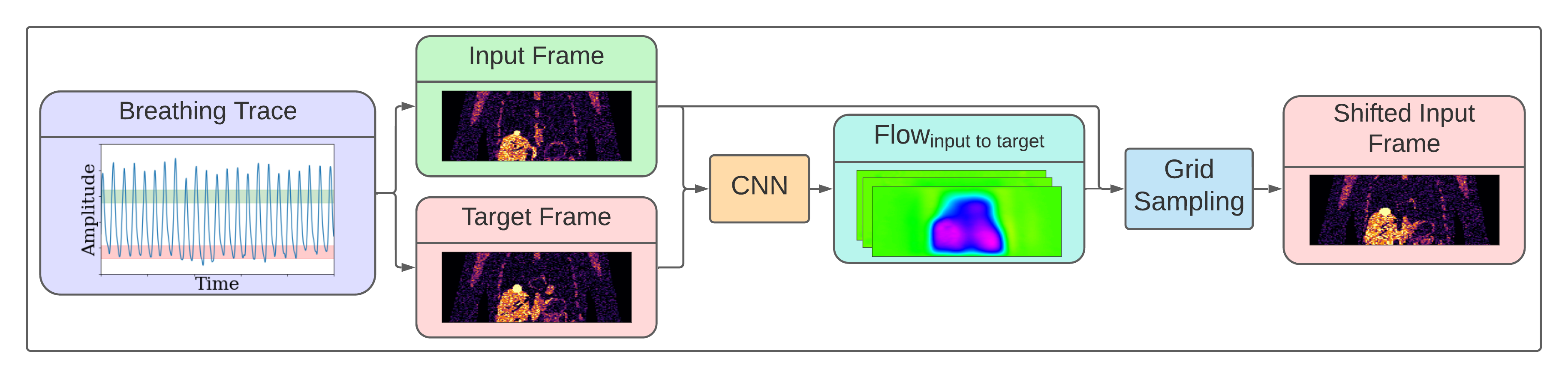}
\caption{A schematic diagram of how a single PET frame is aligned to another frame using \fnpp.
\label{fig1}}
\end{figure*}

%% file: sections/3_Experiments.tex
\section{Experiments}
\label{sec:experiments}

\subsection{Model Selection}
\label{sec:experiments_selection}

Throughout training, not only were the two loss terms evaluated on both the training and validation sets, but the accuracy of the optical flows was evaluated for the validation set using a median-absolute-error metric. This later evaluation was not used for training; it helped assess the true performance of the model both during training and after the training had finished. Having this robust metric allowed for the selection of network and training parameters such as the model architecture, learning rate, loss weights, number of training iterations, and image smoothing parameters. While these tests were too extensive to include in this paper, it is worth mentioning that these parameters were optimized by comparing the predicted optical flows against the ground truth optical flows within the validation set.  The training progress of the final model is shown in \figref{fig12} in \appenref{appendix_training}.

\subsection{Testing on XCAT Frames}
\label{sec:experiments_xcat}

The network was first applied to a test set resembling the same format of the data used to train the network. This test set consisted of six phantoms that were created with identical parameters except for the extent of breathing motion, where the maximum diaphragm motion was varied from 9 mm to 21 mm (the same range as the training data). Each phantom had ten PET frames spaced across a single breathing cycle and a lung lesion with a diameter of 25 mm was placed at the top edge of the diaphragm where the largest amount of motion takes place. Additionally, there were 12$\times$10$^6$ counts spread out equally across the ten frames using the same sampling method as the training set. The accuracy of the optical flow estimates was assessed by predicting the flow from peak inhalation to the end of exhalation and comparing this prediction to the ground truth flow, only considering pixels where activity was present.

Using this same test set, nine of the ten frames from each phantom were shifted using \fnp to be aligned with the tenth frame and then summed together. This provided a “corrected image” that was compared to an image produced without motion. The improvements were quantified by creating a mask that selected a volume of interest (VOI) containing the tumor pixels in the no-motion, uncorrected, and corrected images. Using these masks, three metrics were computed: (1) the intersection over union (IoU) of each mask with the no-motion mask, (2) the activity enclosed in each image using the no-motion mask, and (3) the coefficient of variation (CoV) in each image using the no-motion mask. The IoU between two VOIs is defined as the ratio of the volume of overlap to the volume of the combined VOIs. The CoV is the ratio of the standard deviation to the mean within the VOI. The relative improvements for each of these three metrics were calculated as the percentage of the original residual in the metric (between the no-motion image and the non-corrected image) that was resolved with the \fnp corrections.

\input{figures/fig2}

\subsection{Comparisons to Retrospective Phase Binning}
\label{sec:experiments_rpb_compare}

To compare the improvements against those achieved with retrospective phase binning (RPB) \citep{Didierlaurent2012}, another test set was generated to introduce motion within each bin. For this, the breathing amplitude was varied within the same range as before, but it was divided over 24 time frames rather than ten. Additionally, the PET data collection was modeled by using a clinical breathing trace and the number of counts in each frame was proportional to the time spent in that frame. To create the RPB image, the breathing pattern was split into six equally-spaced phase bins and the detected events in the most quiescent phase were selected as the RPB image. To produce the \fnpp-corrected image, the breathing pattern was split into ten equally-sized amplitude bins between the minimum and maximum extent of the breathing motion. Amplitude binning was chosen to limit the motion within each bin \citep{olsen2008, abdelnour2007, dawood2007}. The difference between binning the time points based on the phase and amplitude is visualized in \figref{fig2}. To easily compare against RPB, the \fnpp-corrected image was created by shifting nine of these ten bins to align with the bin that had the largest overlap with the quiescent phase. Finally, a separate image without motion was created in the quiescent phase. Each of these images (uncorrected, RPB, \fnpp-corrected, and no motion) had a total of 12$\times$10$^6$ counts; however, it is important to note that, in theory, the RPB image required a scan duration that was six times longer in order to “acquire” the same number of counts found in the other images. As before, the IoU, number of enclosed counts, and the CoV of the tumors in each image were compared.

\subsection{Comparisons using Monte Carlo Data}
\label{sec:experiments_mc}

Monte Carlo (MC) simulations of PET imaging were used to generate images whose characteristics more accurately emulated those of clinical patient images. These simulations were accomplished using our previously-validated pipeline of a Siemens Biograph mCT scanner \citep{obriain2022}. The RPB, \fnpp-corrected, and no-motion images were simulated to involve 100 seconds of scan time; however, the RPB required a total scan duration of 600 seconds to acquire the same amount of signal. To obtain images that were more representative of the raw counts, the images were reconstructed without decay correction and no calibration factor was applied. As a result, we used the images in units of counts instead of activity concentrations. Six phase binned attenuation maps were generated for the attenuation correction and the phase with the largest overlap was used for the attenuation correction of each PET image. 

The MC simulations helped determine if the discrepancies found between the RPB method and the corrections provided by \fnp would be mitigated by the blurring due to the actual PET imaging procedure. Through the simulation process, the exact locations of the pixels corresponding to the tumor activity were lost; therefore, these images were used for a qualitative comparison between the two methods.

%% file: figures/fig2.tex
\begin{figure}[t]
\centering 
\includegraphics[width=\columnwidth]{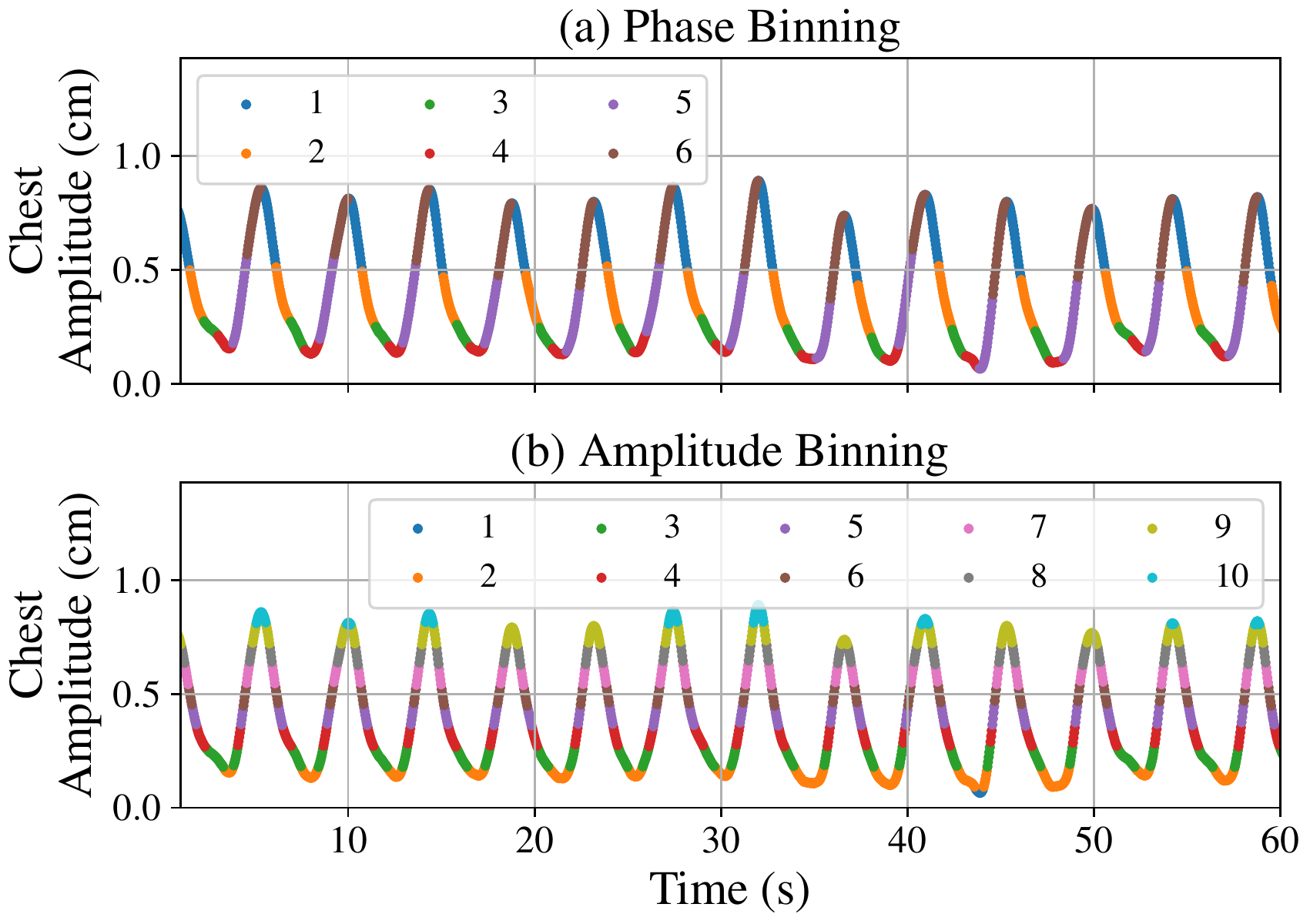}
\caption{Time points in a clinical breathing trace binned based on (a) phase and (b) amplitude.
\label{fig2}}
\end{figure}

%% file: sections/4_Results.tex
\section{Results}
\input{figures/fig3}

\label{sec:results}

\subsection{Testing on XCAT Frames}
\label{sec:results_xcat}

An example of two XCAT frames (full inhalation and full exhalation) from the test phantom with a maximum diaphragm shift of 15 mm is shown in \figref{fig3}. The predicted and ground truth optical flows between the two frames are also presented, showing some over-smoothing present in the \fnp prediction. The breathing motion also resulted in shifts in the lateral and anterior-posterior directions (not shown) that caused the tumor to move in and out of the coronal views. The absolute residual errors in the optical flow predictions are summarized in \figref{fig4}. In all cases, the median absolute residual error was found to be smaller than the slice thickness (2 mm) and the pixel width (4 mm) of the images. In fact, 75th percentile of the residual errors were all within the pixel width of 4 mm.

\input{figures/fig4}

\input{figures/fig5}
\input{figures/fig6}

\figref{fig5} shows the sum of all ten frames from the test phantom with a maximum diaphragm shift of 21 mm before and after the \fnp correction was applied. A summary of the correction performance is shown in \figref{fig6} along with the improvements relative to the uncorrected images. Across the range of tested breathing amplitudes, the average relative improvements were 64\%, 89\%, and 75\% for the IoU, total activity, and CoV, respectively.

\input{figures/fig7}
\input{figures/fig8}
\input{figures/fig9}

\subsection{Comparisons to Retrospective Phase Binning}
\label{sec:results_rpb_compare}

The top row of \figref{fig7} shows a visual comparison of the improvements provided by RPB and \fnp when applied to the binned XCAT PET frames with a diaphragm shift of 18 mm. As shown in \figref{fig8}, both methods provide a similar correction performance. For instance, the average relative improvements with \fnp vs. RPB were IoU: 40\% vs. 44\%, total activity: 83\% vs. 85\%, and CoV: 71\% vs. 75\%, respectively. However, RPB involved a scan time that was six times longer than that of the image corrected with \fnpp.


\subsection{Comparisons using Monte Carlo Data}
\label{sec:results_mc}

The bottom row of \figref{fig7} shows a visual comparison of the improvements provided by RPB and \fnp when applied to the MC-simulated XCAT phantom with a diaphragm shift of 18 mm. Both methods are found to largely mitigate the breathing motion artifacts. As a qualitative assessment, \figref{fig9} shows the profiles drawn through the center of the tumor. As expected, the counts in both the RPB and \fnp profiles are more condensed compared to the uncorrected image, creating tumor profiles with a higher peak.

%% file: figures/fig3.tex
\begin{figure}[t]
\centering 
\includegraphics[width=\columnwidth]{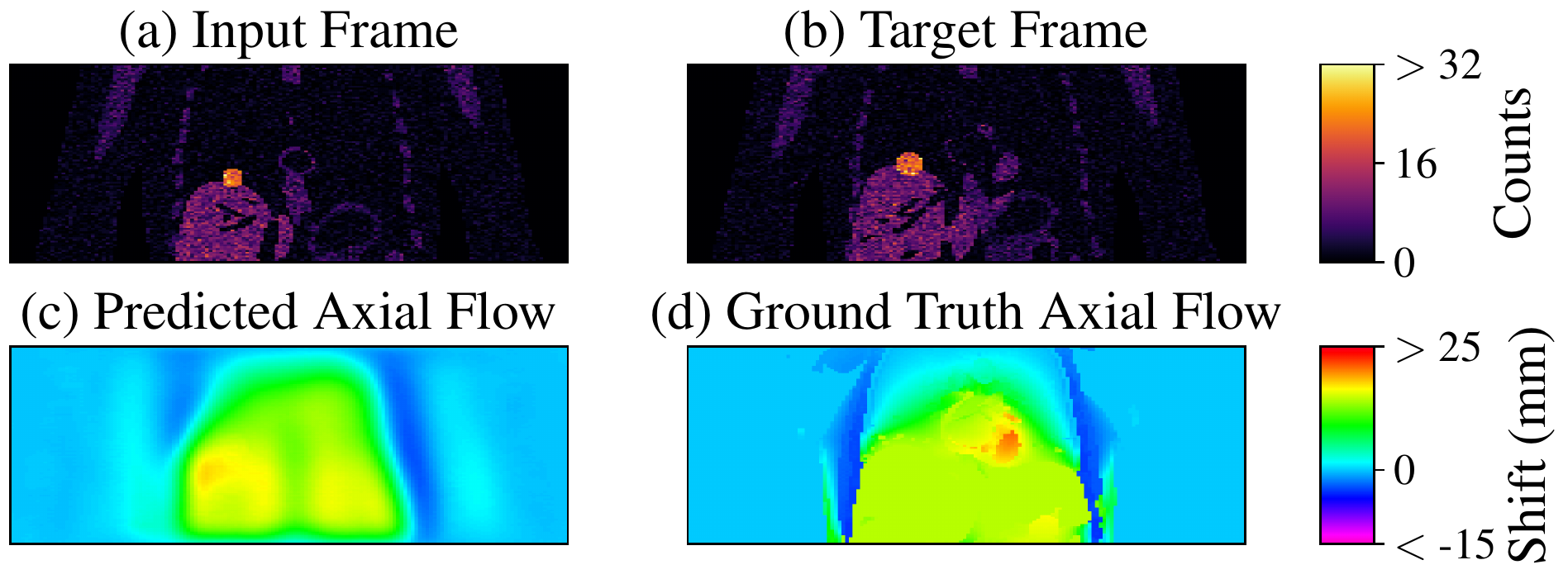}
\caption{Coronal views of (a \& b) two XCAT frames from the same phantom, (c) the predicted axial optical flow, and (d) the ground truth flow between the two frames.
\label{fig3}}
\end{figure}

%% file: figures/fig4.tex
\begin{figure}[t]
\centering 
\includegraphics[width=\columnwidth]{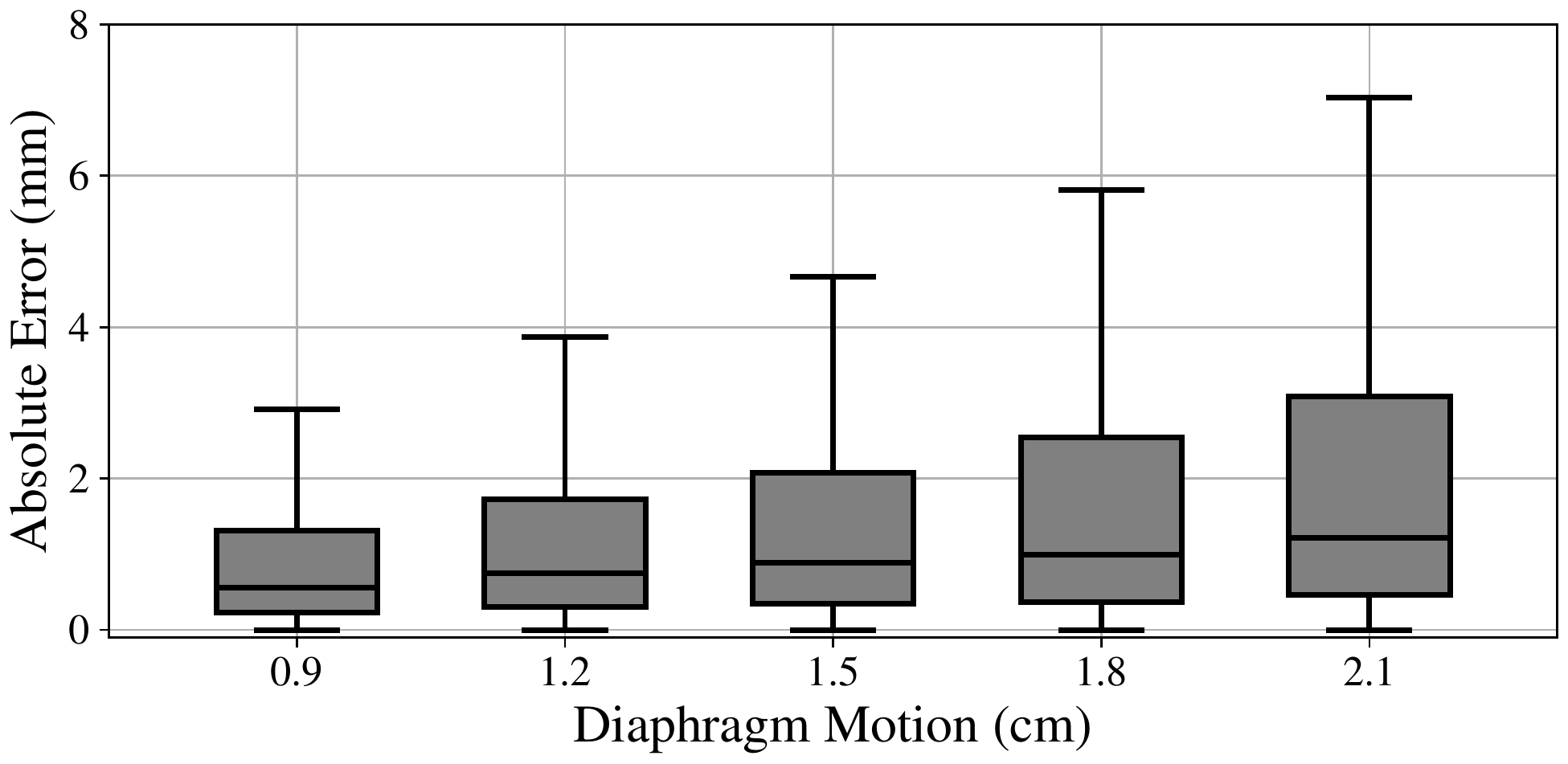}
\caption{Distributions of the absolute residual errors between the predicted and true optical flows from the maximum inhale frame to maximum exhale for a variety of simulated motion extents. Each box extends from the lower to upper quartile values of the data, with a line at the median. The whiskers show the range of $1.5\times$ the interquartile range beyond the lower and upper quartiles.
\label{fig4}}
\end{figure}

%% file: figures/fig5.tex
\begin{figure*}[t]
\centering 
\includegraphics[width=\linewidth]{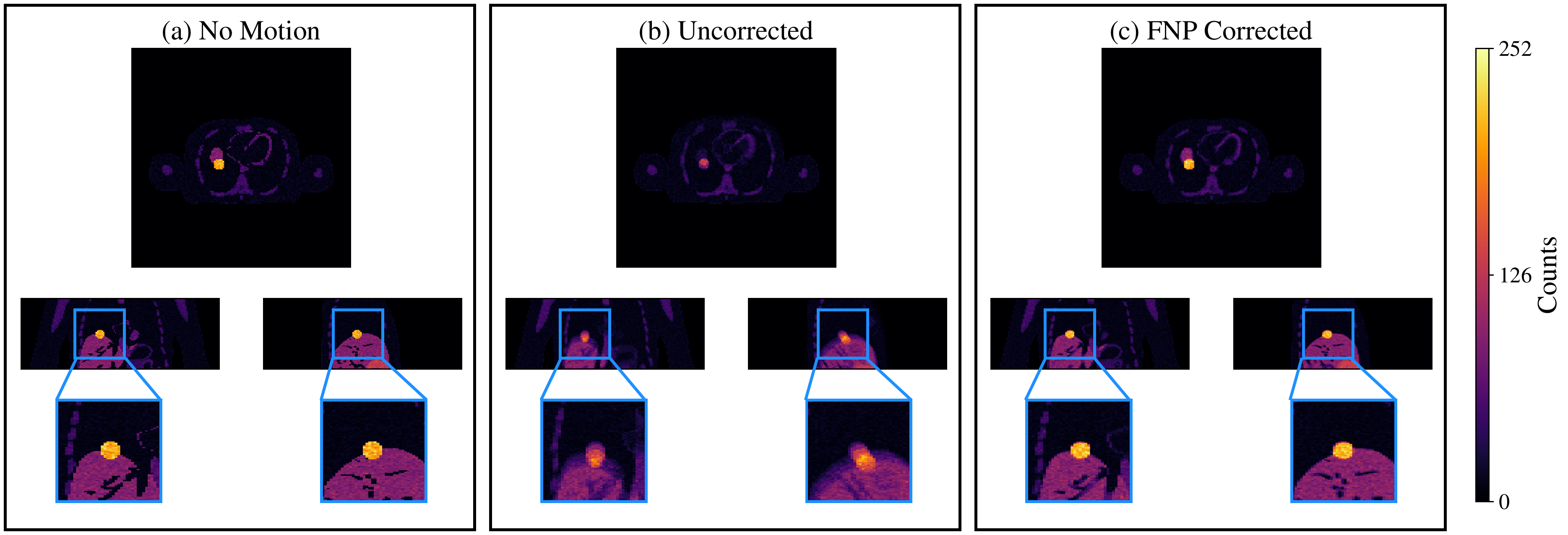}
\caption{An example of an XCAT PET image with (a) no motion, (b) motion, and (c) the motion correction accomplished by \fnpp.
\label{fig5}}
\end{figure*}

%% file: figures/fig6.tex
\afterpage{\begin{figure}[h]
\centering 
\includegraphics[width=\columnwidth]{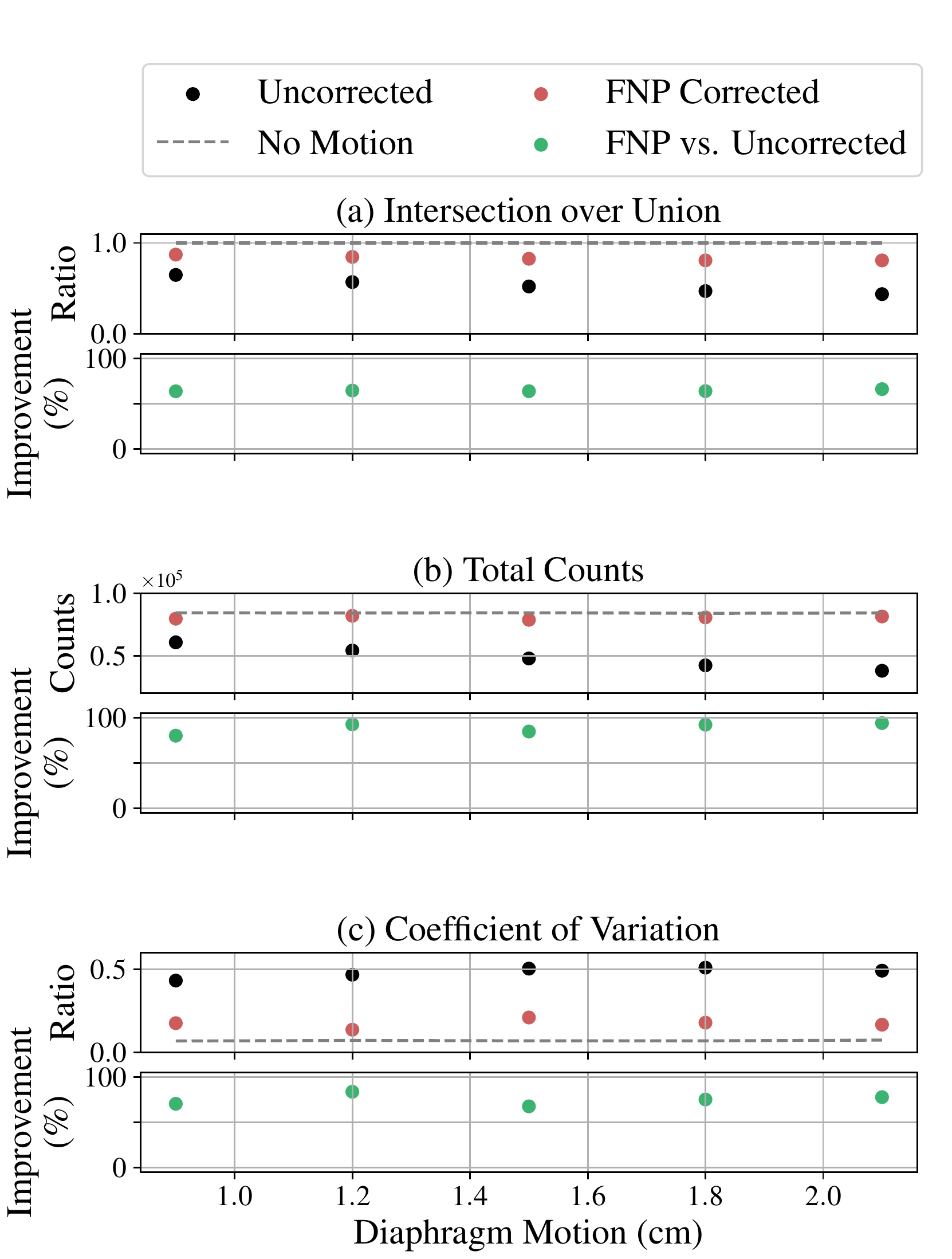}
\caption{Image comparisons with the no motion image in terms of the (a) intersection over union, (b) total number of enclosed counts, and (c) coefficient of variation both before and after the \fnp (FNP) corrections were applied. The relative improvements are shown for a range of breathing motion amplitudes.
\label{fig6}}
\end{figure}}

%% file: figures/fig7.tex
\afterpage{\begin{figure}[t]
\centering 
\includegraphics[width=\columnwidth]{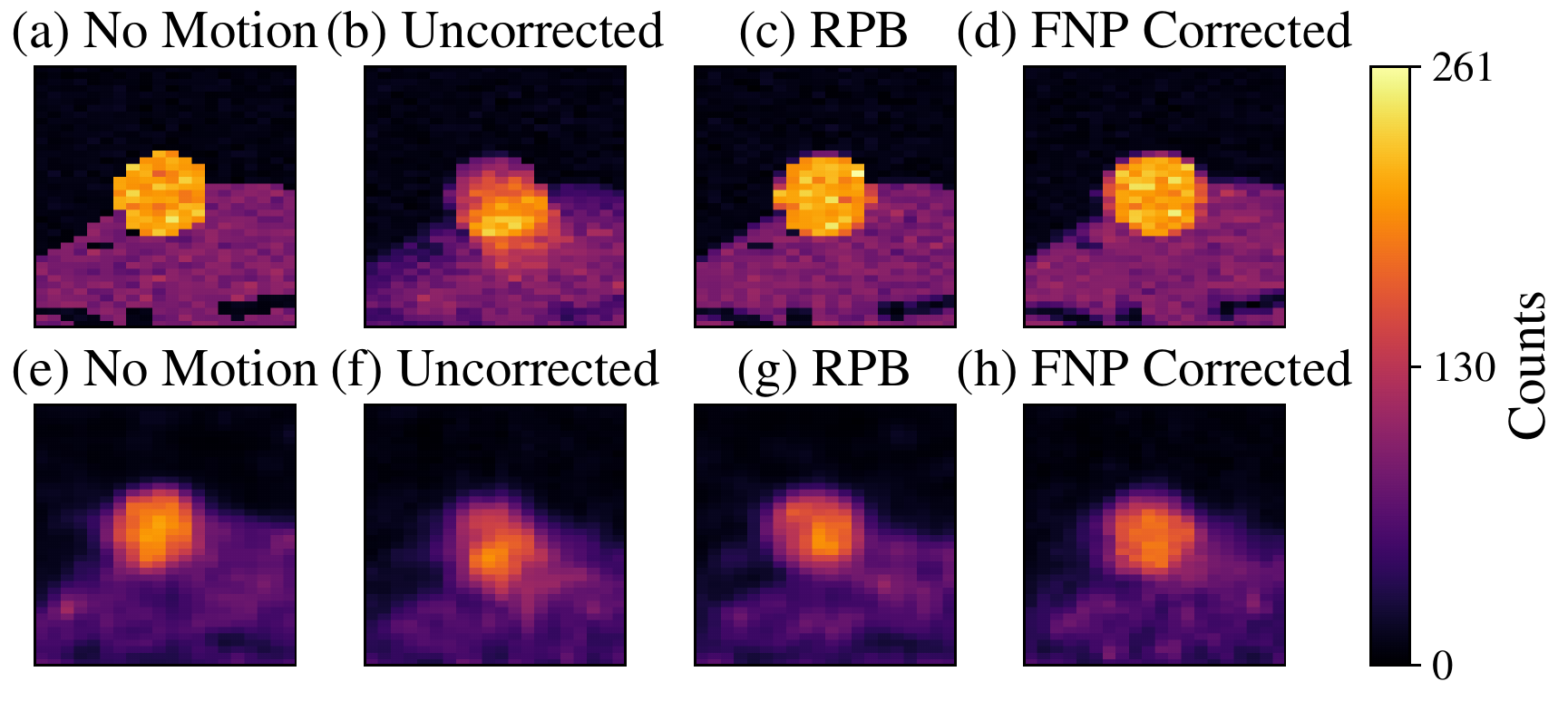}
\caption{Sagittal views of the tumor in the XCAT phantom (a \& e) without motion, (b \& f) with motion, (c \& g) retrospectively phase binned, and (d \& h) with motion but corrected for with \fnpp. The top row of images depicts the results when applied to the binned XCAT phantoms, whereas the bottom row is applied to Monte Carlo data.
\label{fig7}}
\end{figure}}

%% file: figures/fig8.tex
\afterpage{\begin{figure}[t]
\centering 
\includegraphics[width=\columnwidth]{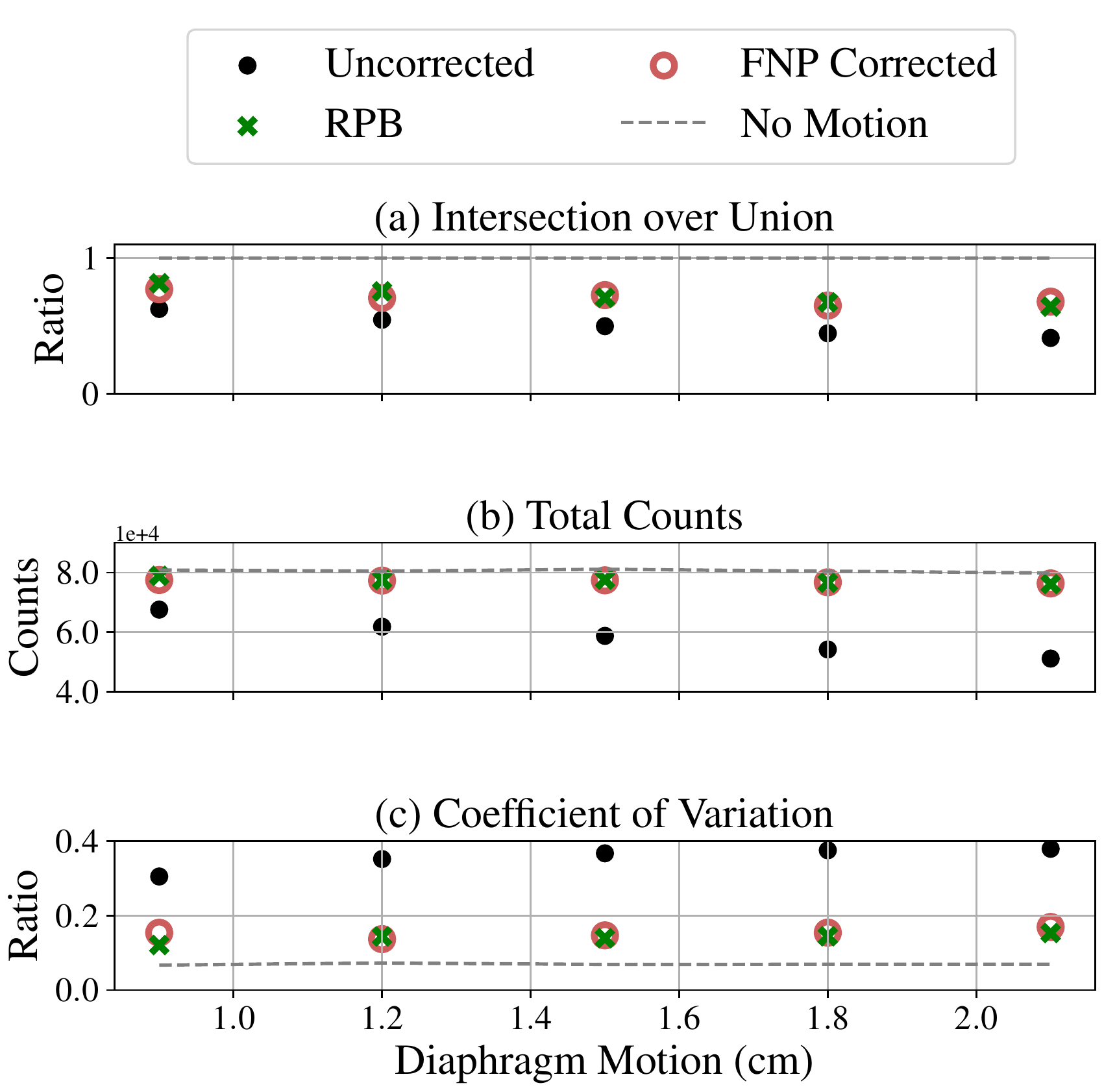}
\caption{Motion correction metrics comparing retrospective phase binning (RPB) to the corrections provided by \fnp (FNP).
\label{fig8}}
\end{figure}}

%% file: figures/fig9.tex
\afterpage{\begin{figure}[t]
\centering 
\includegraphics[width=\columnwidth]{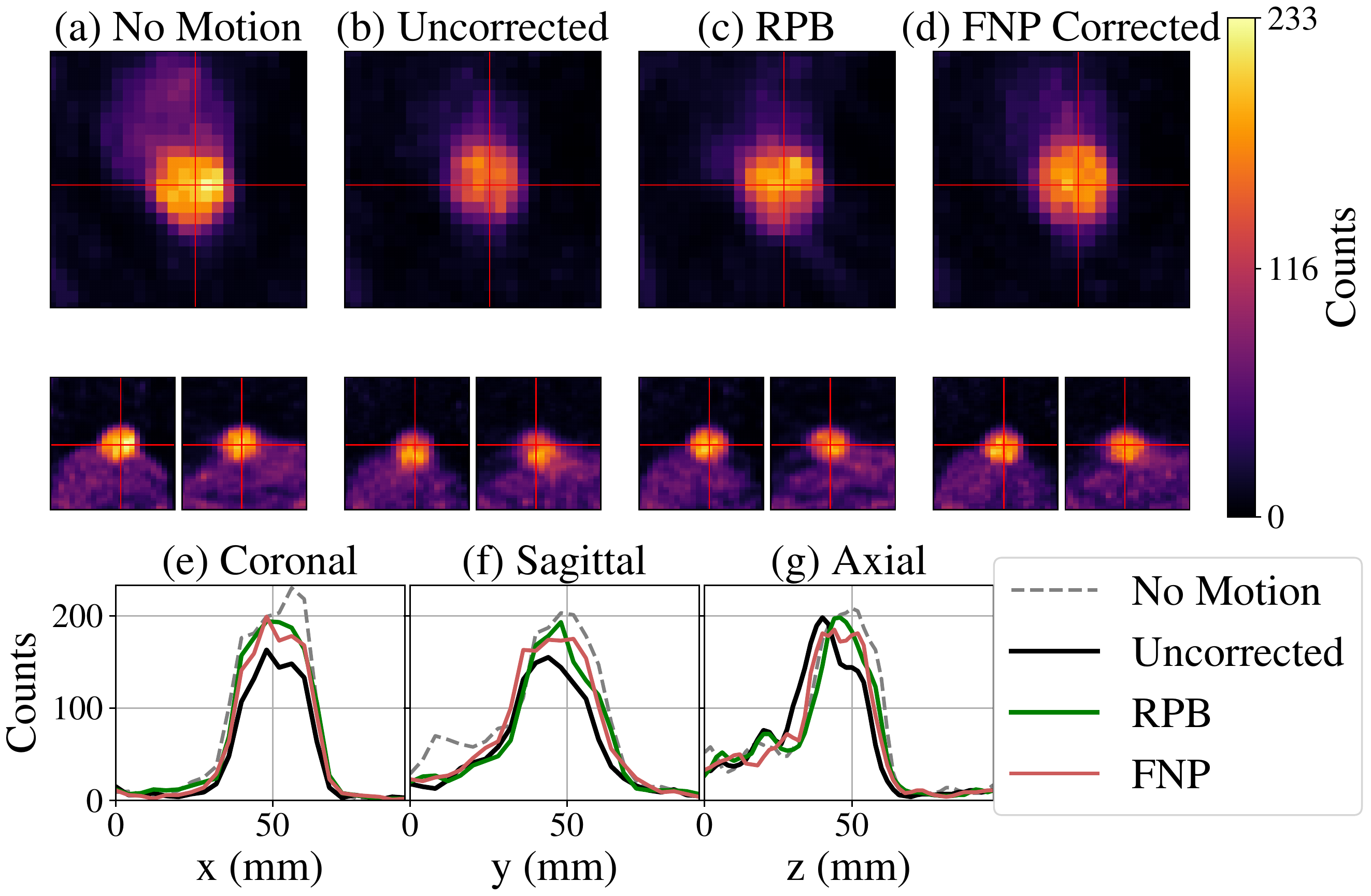}
\caption{Profiles drawn through the center of the tumor in the Monte Carlo-generated PET images (a) with no motion, (b) with motion, (c) retrospectively phase binned, and (d) with motion but corrected with \fnpp.
\label{fig9}}
\end{figure}}

%% file: sections/5_Discussion.tex
\section{Discussion}
\label{sec:discussion}

In PET imaging, the pixel intensities are related to the number of counts detected by a particular pair of detectors. Therefore, it is reasonable to assume that – when shifting the input frame using the grid sampler – individual counts should not be removed or multiplied. This type of constraint is not enforced in the traditional FlowNet framework, since the typical datasets are constructed from images of pixelated color intensities. However, to constrain the network to not artificially add or subtract detections in \fnpp, the invertibility loss was optimized during training.

Additionally, in previous applications of FlowNet frameworks, a smoothness loss term has been used to constrain the network to produce optical flows that are spatially consistent \citep{yu2016}. This loss term was attempted in \fnpp, however, the performance was found to degrade. Interestingly, the effect of producing smooth optical flows was still accomplished through the use of the invertibility constraint, and as a result, a smoothness loss term was not included in our work. Evaluations like this and the choices outlined in \secref{experiments_selection} are only possible due to the access to ground truth optical flows. Therefore, developing \fnp with XCAT data is ideal and these choices would be transferred to the model applied to clinical data.

To perform the motion correction with \fnpp, the PET data was binned based on the amplitude of the breathing motion. Amplitude binning was chosen to limit the amount of motion within each bin, which is visualized in \figref{fig2}. It can be seen that amplitude binning results in a similar amount of motion within each bin, whereas phase binning results in less motion in some bins and more motion in others. As a result, when adopting RPB, it is logical to use phase binning since only the detections acquired within a single bin (with a small amount of motion) are selected and the remainder of the data is disregarded. It is also evident in \figref{fig2} that a few of the amplitude bins account for a small fraction of the total scan time. This is not an issue for \fnp because these bins will only have a small number of counts in their images and have a minimal contribution to the overall corrected image. The choice of using ten amplitude bins was made to provide the majority of the bins with a high enough signal-to-noise yet still limit the amount of motion within each bin. However, if \fnp was used in a clinical setting, the number of bins could be left as a parameter to be chosen by the clinician since this does not require retraining the model.

An example of a predicted optical flow was compared against the ground truth flow in \figref{fig3} to exemplify a limitation of this method. Namely, \fnp is found to produce optical flows that are over-smoothed, which is a characteristic of the FlowNet architecture \citep{Savian2020} and is likely exaggerated by using an unsupervised objective. For instance, if the network was trained with supervised learning – comparing the predictions to the ground truth flows - the optical flows that \fnp produced would likely be much closer to the ground truths. However, performing such supervised learning with clinical data would be very challenging, if not impossible.

As seen in \figref{fig5}, the inaccuracies in the optical flow estimates do not translate into noticeable errors in the corrected images. This is not surprising because the unsupervised learning objective was a comparison between the shifted and target frames in the image-space. Nevertheless, there are a few solutions that could help overcome the limitation of producing over-smoothed optical flows. For example, \fnp could be initially trained with a supervised objective function by using the XCAT phantoms, then fine-tuned with clinical data using an unsupervised learning objective. Another solution would be to adopt a different framework that produces optical flows by iteratively updating the flow matrix \citep{Teed2020}. These iterative methods have been found to produce sharper optical flows compared to frameworks such as FlowNet. 

Another limitation that can be observed in \figref{fig3} is that \fnp does not predict the heart motion accurately. Evidently, since the breathing motion is used to separate the frames in the training set, the network has learned to ignore heart motion because it is not correlated with the phase of the breathing motion. This type of interpretability of the framework potentially enhances the applicability of the method. This is because - rather than having the neural network apply the transformation internally - the model produces an optical flow, which is human interpretable. The optical flow is then applied to the input frame through a grid sampling algorithm that can also be easily understood. In this sense, having a network that produces the optical flows rather than applying the transformations internally is desirable. We leave the door open to training \fnp with cardiac-gated data.
 
As seen in \figref{fig6}, the relative improvements obtained with \fnp were fairly constant across the tested breathing amplitudes. Of the metrics used to evaluate image correction, the enclosed activity is likely the best representation of tumor depiction accuracy. For instance, if a small amount of activity is left outside of the true tumor volume, this could have a large effect on the IoU, yet it will not substantially change the image quality. Conversely, this will only impact the enclosed activity metric when the activity left out of the true volume is substantial.

To introduce some variability in the training set, several XCAT parameters were varied randomly. However, when applying the method to clinical data, the necessary variations would be intrinsic to the training set. Additionally, newly acquired images could be added to the training set to continually update the network. For an unsupervised model such as \fnpp, separating the training and test data may be interesting for a proof-of-concept; however, there is no need to keep the two separate once the method is being applied in a clinical setting. Furthermore, continually fine-tuning the network on new data could be made specific for each institution that adopts the method. In this scenario, the network could be updated based on newly acquired data from that particular clinic; therefore, the network would be more specifically tuned to the protocols and scanners of that institution. 
 
When comparing \fnp to RPB, the performance metrics showed similar results for both methods and the images shown in \figref{fig7} had indistinguishable differences. Furthermore, as seen in \figref{fig7}, the PET spatial resolution limitations were found to be more substantial than any visual differences in the images produced by the two methods. To easily compare the RPB and \fnp images, the amplitude bin that had the largest overlap with the quiescent phase was chosen as the target. However, when adopting \fnp in a clinical setting, it would likely make sense to choose the amplitude bin with the largest number of counts as the target bin.

When applying the network to the MC simulated data, even though the model was not trained on these types of images, \fnp was still effective at reducing the motion artifacts. It can be expected that, when trained on clinical images, the network would be even more effective since it would have learned from images that resulted from the same imaging process.
 
Another intuitive way to register the PET data would be to align the detections before the images are reconstructed using the sinogram data. Applying the corrections in the sinogram-space would have the advantages of not requiring multiple gated low-dose CT images for attenuation corrections and also improving the signal-to-noise of the sinograms before conducting the image reconstructions. One issue that would be encountered when estimating the optical flow between two singorams is due to the discontinuities between the segments within each sinogram. However, a solution to this problem would be to train a separate \fnp model on each segment of the sinogram data. As a result, the sinograms would first be separated into their segments, each segment would be aligned using the corresponding \fnp model and the segments would then be recombined to create a shifted sinogram that is aligned with the target sinogram.

It is important to note some of the limitations of this method. Most importantly, \fnp is capable of correcting for the motion between the different amplitude bins, but cannot correct the motion within the individual bins. Therefore, the method is limited by the gating method that is adopted. In addition to not correcting for cardiac motion, any movement of the patient within the scanner will not be corrected for properly. In this scenario, the movement would result in the blurring within one or more of the binned images and this blurring would remain present after the corrections were applied. Therefore, when applied to clinical data, it would likely be beneficial to have individual time points flagged as unusable when these types of circumstances arise. Furthermore, \fnp is limited by the signal-to-noise ratio of the individual PET frames; having fewer counts in the individual frames will result in poorer predictions. Accordingly, using the frame with the largest number of counts as the target bin would provide the best starting point to perform the corrections.

The applicability of this method is not limited to PET; the \fnp framework could be applied to low-dose CT images as well as other imaging modalities. Furthermore, it could be used dynamically during radiation therapy. For instance, amplitude-binned low-dose CT images could be acquired prior to treatment to predict the optical flow between the different bins. Then, during treatment, only the patient's breathing would be monitored in order to determine the current and next bin. Using the optical flow between these two bins, the treatment volume could be traced to its new location within the patient. Since the optical flow predictions were made prior to the treatment, this process would be efficient and could be applied in real-time.

The code required to train \fnpp, the analysis code, and datasets have been made publicly available (https://github.com/teaghan/FlowNet\_PET).

%% file: sections/6_Conclusion.tex
\section{Conclusion}
\label{sec:conclusion}

\fnp was developed to correct for breathing motion in PET imaging using unsupervised learning. This framework was applied to XCAT phantom data to illustrate a proof-of-concept. \fnp produced interpretable optical flows that were used to shift the amplitude-gated images into a single bin. The mitigation of breathing motion artifacts was shown both qualitatively and quantitatively. Lastly, when compared to the retrospective phase binning method, \fnp was found to provide similar results, but only required one sixth of the scan duration. Since the training is unsupervised, the method is transferable to clinical data without the need for human intervention.

%% file: sections/7_Appendix.tex
\appendix

\section{The Network Architecture}
\label{sec:appendix_architecture}

\input{figures/fig10}

A schematic diagram of our CNN is shown in \figref{fig10}. The network architecture was based on “FlowNet Simple” \citep{Fischer2015} where the 2D convolutions were replaced with 3D ones to account for PET images. Transitioning the network from a 2D-CNN to a 3D-CNN comes with substantial costs in memory use and computation time. Therefore, the network was simplified to consist of fewer convolutional layers and filters, both of which can be inferred from \figref{fig10}. As seen in this Figure, the network takes two PET frames as inputs and subsequently downsamples the images through a series of 3D convolutions. Each of the downsampling blocks consists of two 3D convolutional layers; both layers use a filter size of (3$\times$3$\times$3) pixels, but the first layer uses a stride length of two pixels in all three directions whereas the second layer uses a stride length of one pixel. The convolutions are each followed by a ReLU activation. Following the downsampling blocks, the feature map is then upsampled in stages, where each upsampling operation includes a single transposed convolutional layer with a filter size of (4$\times$4$\times$4) pixels and a stride length of two pixels, which is followed by a ReLU activation. 

At each intermediate resolution, there is a flow estimator, which consists of a single convolutional layer with a filter size of (5$\times$5$\times$5) pixels and a stride length of one pixel. Each flow estimator receives the intermediate outputs produced throughout the network that are at the matching resolution, including the upsampled lower resolution flow. The combination of these inputs is clearer in \figref{fig10}. The idea behind these skip connections is to provide the later parts of the network with information that has potentially been lost in previous operations and make the learning process easier \citep{He2016}. The final outputs of the network are the optical flows estimated at each resolution, including the original resolution of the input frame. This allows for comparisons to be made between the shifted input frame and target frame at various resolutions, which also helps provide a smoother learning process \citep{Fischer2015}.

\section{Grid Sampling}
\label{sec:appendix_gridsample}

\input{figures/fig11}
\setcounter{figure}{0}

Understanding the training objectives outlined below requires a brief background on the grid sampling procedure, $T$, that performs the transformation to align the input frame, $I_{inp}$, with the target frame, $I_{tgt}$. An example of how this operation works for a small 2D grayscale image is shown in \figref{fig11}. In the case of 3D images, the baseline grid of pixel locations, $G_0$, consists of a 3D grid, where each pixel contains a unique combination of $x,\ y,\ z$ pixel locations. The forward flow from the input to the target frame, $u_{fwd}$, is used to adjust the pixel locations in the grid to produce a new grid, $G_{fwd}=G_0+u_{fwd}$. The transformation is then performed by sampling $I_{inp}$ using $G_{fwd}$ :

\begin{equation}
    I_{shift} =\
    T\left(
    I_{inp},\ 
    G_{fwd}
    \right)\;.
\label{eq:grid_sample}
\end{equation}

\section{Training the Model}
\label{sec:appendix_training}

The objective function used to train \fnp consists of two loss terms: the photometric loss and invertibility loss. For the photometric loss, a Charbonnier penalty function was chosen to mitigate the effects of outliers \citep{sun2013}. Namely, when comparing a shifted frame to its target frame, this loss term can be summarized as

\begin{equation}
    \rho_{photo}(u_{fwd}) =\
    \frac{1}{N}\sum_{i=0}^N\
    \left(\left(I_{shift}^i - \
    I_{tgt}^i\right)^2 \
    + \epsilon^2
    \right)^{\alpha}\;,
\label{eq:photo_loss}
\end{equation}

for N pixels in each frame. To produce the best results, $\alpha$ and $\epsilon$ were chosen to be 0.3 and $1\times 10^{-9}$, respectively.

Counter to the photometric loss term, the invertibility constraint operates on the pixel grid rather than the images themselves. In more detail, the invertibility constraint is enforced by also predicting the backward flow from the target to the input frame, $u_{bwd}$. Then, by using $G_{bwd}=G_0+u_{bwd}$ in the grid sampling procedure to sample $G_{fwd}$, a cycled grid is produced:

\begin{equation}
    G_{cyc} =\
    T\left(
    G_{fwb},\ 
    G_{bwd}
    \right)\;.
\label{eq:cycle_grid}
\end{equation}

\input{figures/fig12}

Accordingly, if $G_{cyc}$  is constrained to return to $G_0$, then $u_{fwd}$ is invertible and $u_{bwd}$ is its inverse.
This constraint is enforced by adopting the invertibility loss term, which can be formalized as

\begin{equation}
    \rho_{inv}(u_{fwd},\ u_{bwd}) =\
    \frac{1}{3N}\sum_{i=0}^N\
    \sum_{j=1}^3\
    \left(G_{cyc}^{i,j} - \
    G_{0}^{i,j}\right)^2 \;,
\label{eq:inv_loss}
\end{equation}

where $j=1,2,3$ represent the $x,\ y,\ z$ coordinates.

Other implementations of FlowNet have utilized a similar cycle-consistency where the constraint is imposed on the cycled input frame \citep{Meister2017}. However, this does not guarantee a one-to-one pixel correspondence since there could be multiple pixels with the same intensity, allowing cycle-consistency to be achieved without ensuring that the flow itself is invertible. With our method, since the baseline grid has a unique combination of $x,\ y,\ z$ locations at every pixel, the only possible way for the constraint of $G_{cyc}=G_0$ to be met is if  $u_{fwd}$ is invertible and $u_{bwd}$ is its inverse

The final loss function that was minimized through training was the sum of the photometric and the invertibility loss terms

\begin{equation}
    \loss (u_{fwd},\ u_{bwd}) =\ \rho_{photo}(u_{fwd}) +\
    \lambda\rho_{inv}(u_{fwd},\ u_{bwd})\;,
\label{eq:total_loss}
\end{equation}

where a weighting factor of $\lambda=1100$ was multiplied to the invertibility term, which helped balance the scaling of the gradients of each loss term. This objective was optimized using the Adam optimizer \citep{Kingma2014} with an initial learning rate of 0.0003, which was decayed by a factor of 0.7 every 8000 iterations. The model was trained for a total of $1.5\times 10^5$ iterations, using one sample per iteration. The two loss terms were evaluated on the training and validation sets throughout the training of the model and are shown in \figref{fig12}. Also shown in this plot is the median-absolute-error evaluated between the predicted and ground truth optical flows in the validation set, which was not used for training.

%% file: figures/fig10.tex
\begin{figure*}[ht]
\centering 
\includegraphics[width=0.85\linewidth]{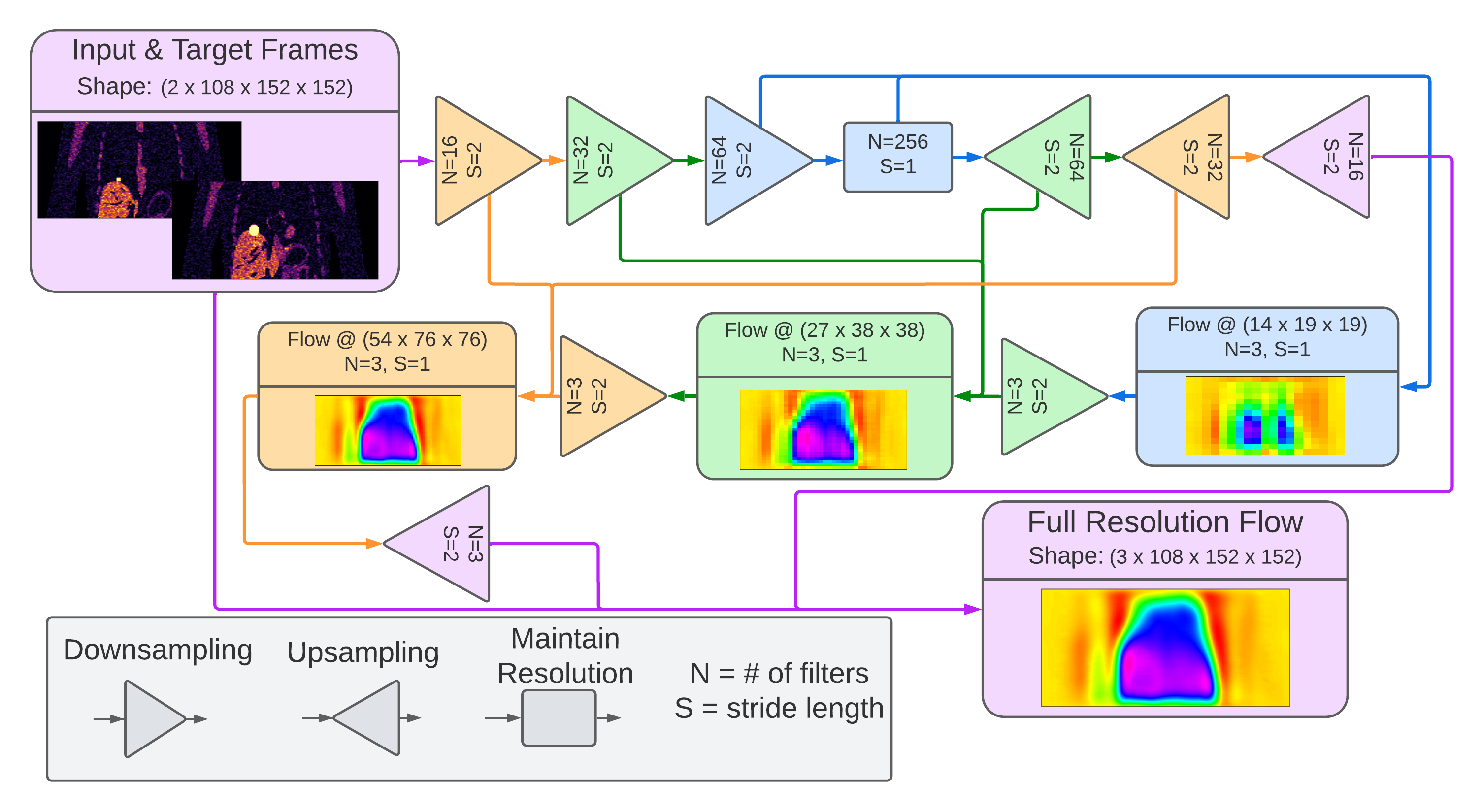}
\caption{A diagram of the CNN architecture used in \fnpp. Each colored shape represents a convolutional block of the network and the arrows represent the connections between them. The color depicts the resolution of the output of that block, whereas the shape describes the type of operation (downsampling, upsampling, or constant resolution). For the blocks that produce outputs with intuitive representations, sagittal slices of the 3D image are shown; the remaining outputs are non-intuitive feature maps.
\label{fig10}}
\end{figure*}

%% file: figures/fig11.tex
\begin{figure}[t]
\centering 
\includegraphics[width=\columnwidth]{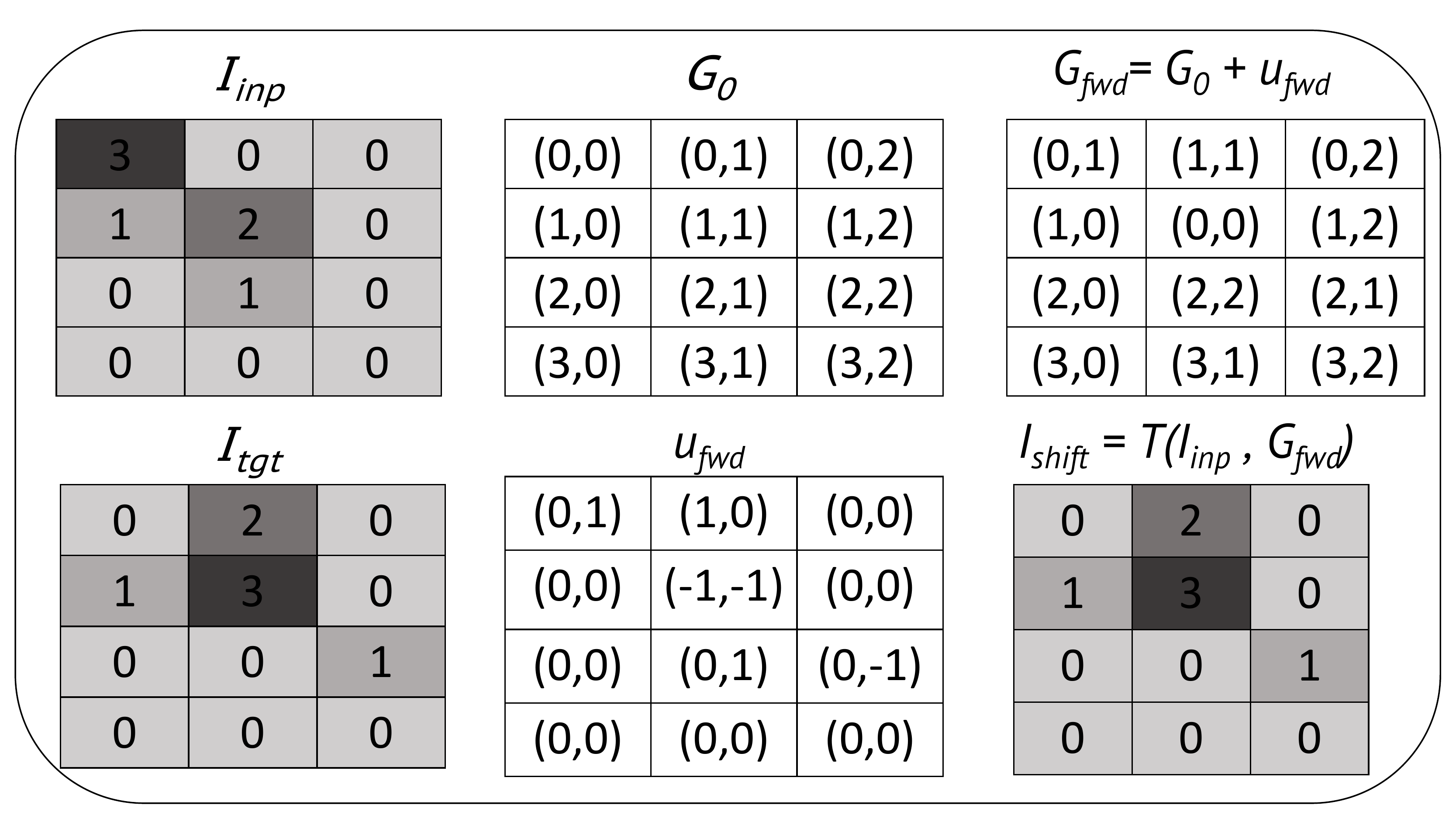}
\caption{An example of the grid sampling procedure for a small 2D grayscale image.
\label{fig11}}
\end{figure}

%% file: figures/fig12.tex
\begin{figure}[ht!]
\centering 
\includegraphics[width=\columnwidth]{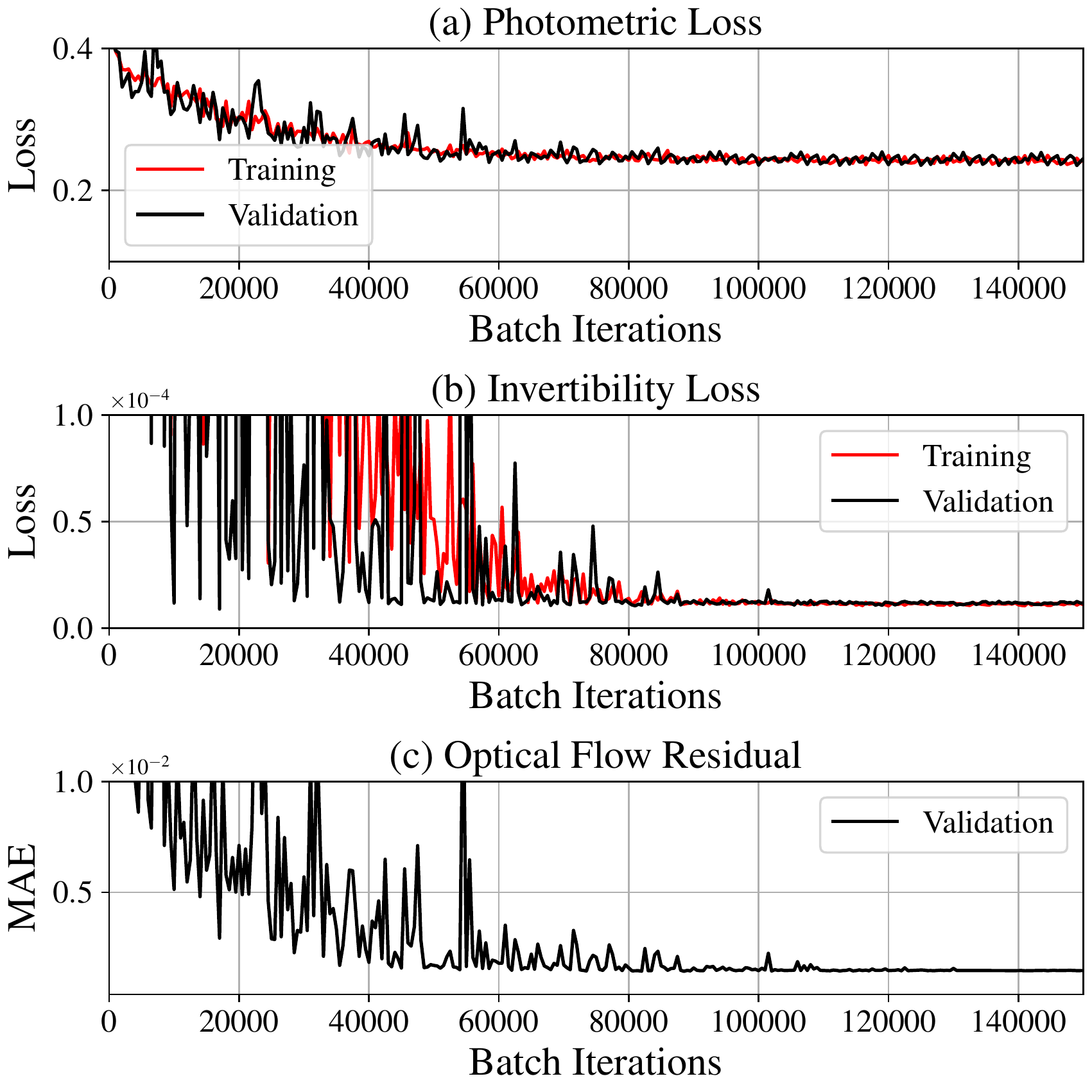}
\caption{The progress of the CNN performance evaluated during training. The (a) photometric loss and (b) invertibility loss were evaluated on the training and validation sets, whereas the (c) median-absolute-error between the predicted and ground truth optical flows were evaluated only on the validation set. This later metric was only used to track the true performance of the network and not for the training.
\label{fig12}}
\end{figure}